\documentstyle[11pt,newpasp,twoside,epsf]{article}
\def\edcomment#1{\iffalse\marginpar{\raggedright\sl#1\/}\else\relax\fi}
\reversemarginpar

\begin{document}
\title{Reddening Independent Quasar Selection from a Wide Field Optical and Near-IR Imaging Survey}
\author{$^{1,2}$Sabbey, C.N., $^{1}$Sharp, R.G., $^{2}$Vivas, A.K., $^{1}$Hodgkin, S.T., $^{2}$Coppi, P.S., $^{1}$McMahon, R.G.}
\affil{$^{1}$Institute of Astronomy, Madingley Road, Cambridge, CB3 0HA, UK}
\affil{$^{2}$Astronomy Department, Yale University, P.O. Box 208101, New Haven, 
CT 06520-8101 USA}

\begin{abstract}
We present preliminary results from a wide field near-IR imaging survey
that uses the Cambridge InfraRed Survey Instrument (CIRSI) on the 2.5m
Isaac Newton Telescope (INT).  CIRSI is a JH-band mosaic imager that
contains 4 Rockwell 1024$^{2}$ HgCdTe detectors (the largest IR camera
in existence), allowing us to survey $\approx 4$ deg$^{2}$ per night to
H $\approx 19$.  Combining CIRSI observations with the deep optical
imaging from the INT Wide Field Survey, we demonstrate a reddening
independent quasar selection technique based on the (g $-$ z) / (z $-$
H) color diagram.
\end{abstract}

\section{Introduction}

To fill the gap between the existing shallow wide-field IR surveys (e.g.,
2MASS, DENIS) and deep surveys over small fields of view, we have begun
a near-IR imaging survey on the 2.5m INT using the CIRSI mosaic imager
(Beckett et al. 1996; Mackay et al.\ 2000).  CIRSI has an instantaneous
field of view of 15.6\arcmin\ $\times$ 15.6\arcmin\ on the INT (0.45
arcsec pixel$^{-1}$), making it ideally suited to large area survey
programs of moderate depth (H = 19--20).  In particular, the survey aims
to cover 50 deg$^{2}$ in the J and H bands with 5 sigma limits of J =
20.6, H = 19.1 ($\sim 3.5$ magnitudes fainter than 2MASS) during a total
of $\approx 30$ nights in the spring of 2000 and 2001.  Even with this
modest investment of telescope time we sample a 'local' volume $\sim
10$\% that of 2MASS.  Thus far we have obtained J and H observations
covering $\approx 5$ deg$^{2}$ in a Southern Galactic Cap equatorial
strip and several 30\arcmin\ $\times$ 30\arcmin\ mosaic images in the
ELAIS, Lockman Hole, and XMM fields.  The fields were chosen to overlap
datasets at other wavelengths, including the deep optical observations of
the INT Wide Field Survey (McMahon et al. 2000), and the XMM X-ray data.

The CIRSI near-IR survey has several scientific objectives, including
identifying and measuring the space density of cool white dwarfs and
T dwarfs, and selecting quasars and damped Lyman-$\alpha$ absorption
systems at IR wavelengths (see Warren et al.\ 1999).  The predicted
surface density of QSOs with $z>2$ and H $< 18.5$ is 5--10 deg$^{-2}$,
and thus in 50 deg$^{2}$ we expect 250--500 QSOs with $z>2$.  The number
density of damped Lyman-$\alpha$ (DLA) systems is dN/dz $\sim$ 0.1
(Storrie-Lombardi, McMahon, and Irwin 1996), and thus 25--50 DLA systems
should be detected.  These DLA systems are important for comparison to
the previous samples detected via optically selected quasars, given the
expected bias of the optically derived samples against lines of sight
that may contain gas rich dusty galaxies (Pei \& Fall 1995).

Here we present the first results from a reddening independent
quasar selection technique based on combined optical and near-IR color
diagrams.  As shown in Figure~1 using model quasar colors, the reddening 
vector in the (g $-$ z) / (z $-$ H) color diagram does not drive quasars 
into the stellar locus (in contrast to purely optical colour-colour plots).
The corresponding observed color diagram for a 30\arcmin\ $\times$ 30\arcmin\
mosaic image in the ELAIS region is shown in Figure~2.  The CIRSI data
were reduced using an automated pipeline (Sabbey et al. 2000), and a
SExtractor object catalog was created and merged with the existing
Wide Field Survey optical catalog in this field to produce the color diagram.
Object identifications from WIYN Hydra follow-up spectroscopy are labelled
in the figure.  Hydra observations were obtained for a total of four
fields during 2 and 3 September 2000, confirming 80 quasars at redshifts
$1 < z < 4$ (including optical-only selected quasars), and reductions and
interpretation are still in progress.  The preliminary result is that
quasar selection using the (g $-$ z) / (z $-$ H) diagram is a viable
technique and is expected to be insensitive to reddening.

\begin{figure}
\plotone{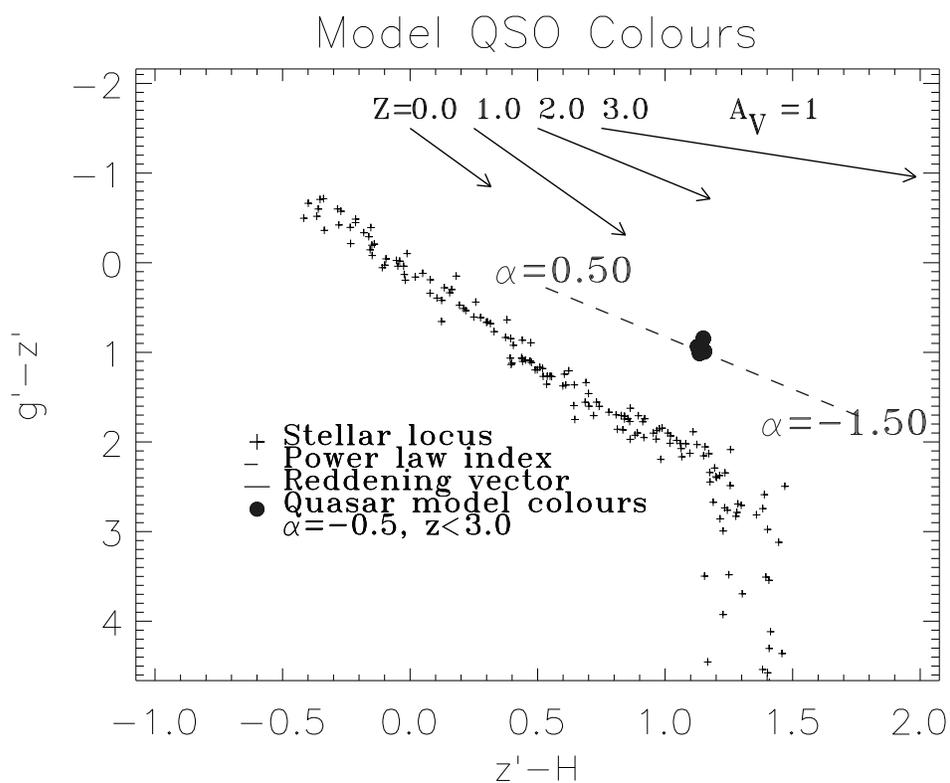}
\caption{A simulated optical and near-IR colour-colour plot showing the locus of quasars compared with Galactic stars from Gunn \& Stryker 1983.  The reddening vector does not drive the quasars into the stellar locus unlike purely optical clour-colour plots.}
\end{figure}

\begin{figure}
\plotone{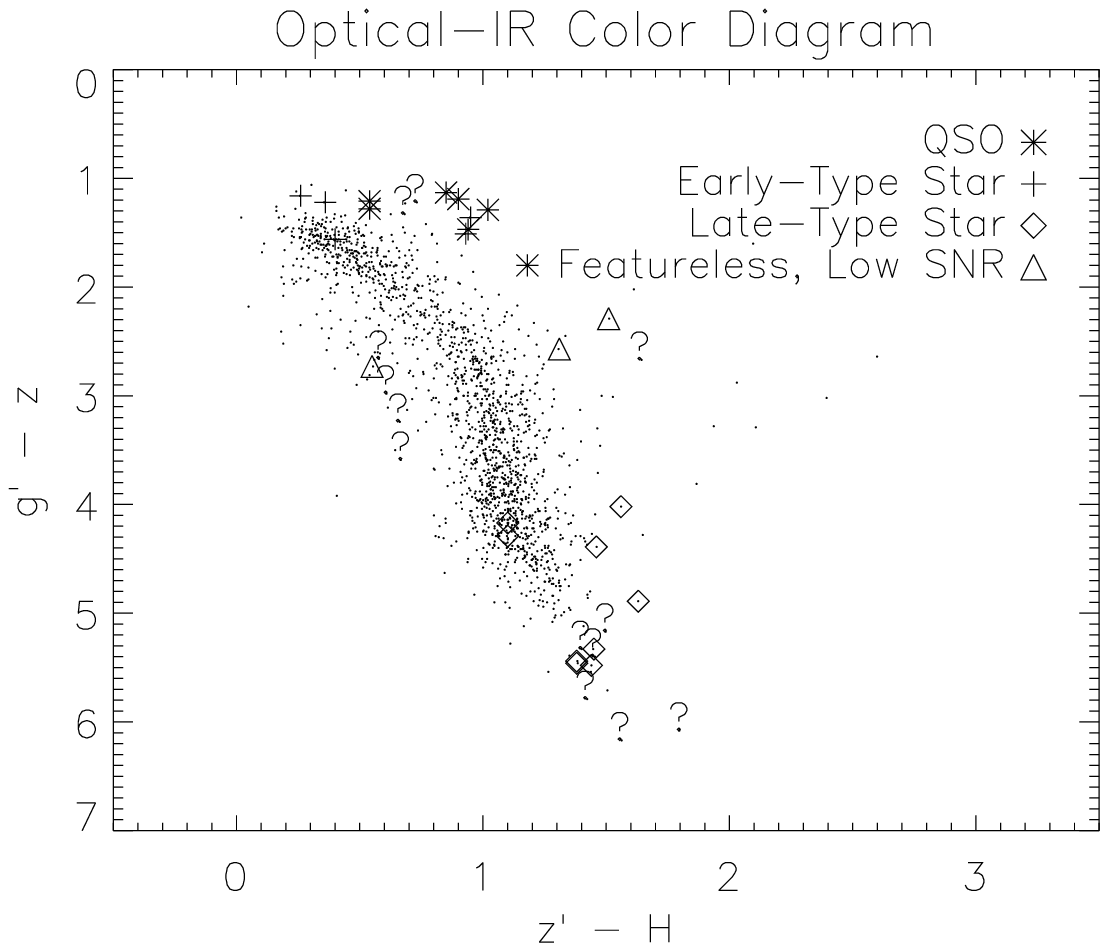}
\caption{The observed optical and near-IR colour diagram using CIRSI and WFS catalogues for a 30\arcmin\ $\times$ 30\arcmin\ field in the ELAIS region.  Object identifications from WIYN Hydra follow-up spectroscopy are labelled.}
\end{figure}


\begin{references}
\reference Beckett, M.G. et al.\ 1996, SPIE, 2871, 1152
\reference Gunn, J.E., \& Stryker, L.L. 1983, ApJS, 52, 121
\reference Mackay, C.D., et al.\ 2000, SPIE 4008, in press
\reference McMahon et al.\ 2000, astro-ph/0001285
\reference Pei, Y.C., \& Fall, S.M. 1995, ApJ, 454, 69
\reference Sabbey et al.\ 2000, ADASS X, ASP Conference Series, in preparation
\reference Storrie-Lombardi, L.J., McMahon, R.G., \& Irwin, M. 1996, MNRAS, 283, L79
\reference Warren, S.J., Hewett, P.C., Foltz, C.B. 1999, MNRAS, 312, 827
\end{references}
\end{document}